# Performance Analysis of Spatial and Transform Filters for Efficient Image Noise Reduction

Santosh Paudel
Computer and Electronics Engineering
KEC, TU
Lalitpur, Nepal
spaudel830@gmail.com

Ajay Kumar Shrestha
Computer and Electronics Engineering
KEC, TU
Lalitpur, Nepal
ajayshrestha1@gmail.com

Pradip Singh Maharjan
Computer and Electronics Engineering
KEC, TU
Lalitpur, Nepal
psmaharjan@kec.edu.np

Rameshwar Rijal
Computer and Electronics Engineering
KEC, TU
Lalitpur, Nepal
rijal_rameshwar@yahoo.com

*Abstract*—During the acquisition of an image from its source, noise always becomes integral part of it. Various algorithms have been used in past to denoise the images. Image denoising still has scope for improvement. Visual information transmitted in the form of digital images has become a considerable method of communication in the modern age, but the image obtained after transmission is often corrupted due to noise. In this paper, we review the existing denoising algorithms such as filtering approach and wavelets based approach, and then perform their comparative study with bilateral filters. We use different noise models to describe additive and multiplicative noise in an image. Based on the samples of degraded pixel neighborhoods as inputs, the output of an efficient filtering approach has shown a better image denoising performance. This yields promising qualitative and quantitative results of the degraded noisy images in terms of Peak Signal to Noise Ratio, Mean Square Error and Universal Quality Identifier.

## I. INTRODUCTION

Noise in an image is a very common problem. An image gets corrupted with noise during acquisition, transmission, storage and retrieval processes. Noise may be classified as substitutive or impulsive noise (e.g., salt and pepper noise, random-valued impulse noise), additive noise (e.g., additive white Gaussian noise) and multiplicative noise (e.g. speckle noise) [1]. The simple median filter works efficiently to suppress impulse noise of low density. However, many denoising schemes have been proposed recently which are efficient in suppressing impulse noise of moderate and high noise densities. In many occasions, noise in digital images is found to be additive in nature with uniform power in the whole bandwidth along with Gaussian probability distribution and is termed as Additive White Gaussian Noise (AWGN). It is difficult to suppress AWGN since it corrupts almost all pixels in an image. The arithmetic mean filter, commonly known as Mean filter, can be employed to suppress AWGN but it introduces a blurring effect [2]. Multiplicative or speckle noise is an inherent property of medical ultrasound imaging. Speckle noise occurs in almost all coherent imaging systems such as laser, acoustics and SAR (Synthetic Aperture Radar) images [3].

This paper introduces different types of noise to be considered in an image and analyzed for various spatial and transforms domain filters by considering the image metrics such as mean square error (MSE), root mean squared error (RMSE), Peak Signal to Noise Ratio (PSNR) and universal quality index (UQI).

## II. BACKGROUND

The Wavelet Transform (WT) is a powerful tool of signal and image processing, which has been successfully used in many scientific fields such as signal processing, image compression, computer graphics, and pattern recognition. WT represents image energy in compact form and representation helps in determining threshold between noisy features and important image feature [4]. The Continuous WT (CWT) technique expands the signal on basis functions created by expanding, shrinking and shifting a single prototype function, which is named as mother wavelet, specially selected for the signal under considerations. This transformation decomposes the signal into different scales with different levels of resolution. Since a scale parameter shrinks or expands the mother wavelet in CWT, the result of the transform appears as time-scale representation. The scale parameter is indirectly related to frequency when the center frequency of mother wavelet is considered. A mother wavelet has a zero mean value, which requires the transformation kernel of the wavelet transform to compactly support localization in time, thereby offering the potential to capture the spikes occurring instantly in a short period of time [2],[5]. A wavelet expansion is a representation of a signal in terms of an orthogonal collection of real-values generated by applying suitable transformation to the original selected wavelets. The main difference between mother wavelet functions such as Haar, Daubechies, Symlets, Coiflets and Bi-orthogonal lies on how their scaling signals and the wavelets are defined.

D. L. Donoho has done a lot of work on filtering of additive Gaussian noise using wavelet soft thresholding [6]. Wavelets play a major role in image compression and image denoising [7]. These Wavelet coefficients calculated by a wavelet transform represent change in the time series at a particular resolution. It is always possible to filter out the noise by considering the time series at various resolutions. The small coefficients are dominated by noise after applying wavelet transform. However, coefficients with a larger absolute value carry more signal information than noise. Replacing the smallest, noisy coefficients by zero and a backward wavelet transform on the result may lead to reconstruction with the essential signal characteristics and reduced noise. For thresholding, there are three observations and assumptions which are given as:

1. The decorrelating property of a wavelet transform creates a sparse signal in which most of the untouched coefficients are zero or close to zero.
2. Noise is spread out equally over all the coefficients.
3. The noise level is not too high, therefore we can recognize the signal and the signal wavelet coefficients.

Thus, the choice of threshold level is an important task. The coefficients having magnitude greater than threshold are considered as signal of interest and the same or modified coefficients are kept according to the selected threshold, whereas other coefficients become zero [8]. Then the image is reconstructed from the modified coefficients. Usually the selection of threshold and the preservation of the edges of the denoised images are important points of interest.

### III. OVERVIEW

This paper basically focuses on the wavelet transform filtering method. All wavelet transform denoising algorithms involve the following three steps in general as shown in Figure 1.

- Forward Wavelet Transform: Wavelet coefficients are obtained by applying the wavelet transform Estimation.
- Clean coefficients are estimated from the noisy ones.
- Inverse Wavelet Transform: A clean image is obtained by applying the inverse wavelet transform.

There are various methods for wavelet thresholding, which rely on the choice of a threshold value. The typically used threshold methods for denoising an image are Visu Shrink, Sure Shrink, Bayes Shrink, Neigh Shrink, Oracle Shrink, Smooth Shrink and Fuzzy based Shrink.

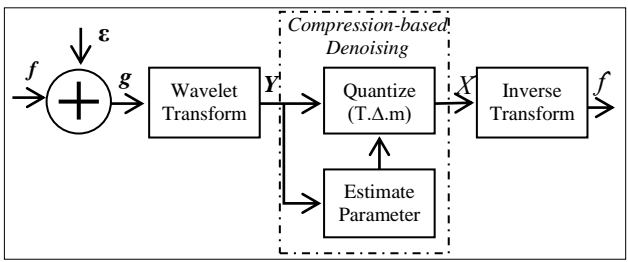

Fig. 1. Denoising using Wavelet Transform.

#### A. Visu Shrink

Visu Shrink is also called as the universal threshold method. The method was introduced by D. L. Donoho et al. [8],[9]. It uses a threshold value *t* that is proportional to the standard deviation of the noise. It follows the hard thresholding rule. It is defined by Eq. (1).

$$t = \sigma\sqrt{2\log n} \quad\quad\quad\quad\quad\quad\quad\quad\quad (1)$$

Here, $\sigma^2$ is the noise variance and n is the number of samples.

An estimate of the noise level $\tilde{\sigma}$ was defined based on the median absolute deviation given by Eq. (2).

$$\tilde{\sigma} = \frac{median\left[\{(g_{j-1,k}): k = 0,1,2\ldots\ldots 2^j - 1\}\right]}{0.625} \quad (2)$$

#### B. Sure Shrink

Sure Shrink is based on Stein's Unbiased Risk Estimator (SURE) and was proposed by Donoho and Johnston [6],[8]. It is a combination of the universal threshold and the SURE threshold. This method specifies a threshold value for each resolution level *j* in the wavelet transform which is referred as level dependent thresholding [10]. The objective of this method is to minimize the mean square error, defined by Eq. (3).

$$MSE = \frac{1}{n}\sum_{x,y=1}^{n}(Z(x_i,y_i) - S(x_i,y_i))^2 \quad\quad (3)$$

Here, $Z(x_i,y_j)$ is the estimate of the signal, $S(x_i,y_j)$ is the original signal without noise and n is the size of the signal. Sure Shrink suppresses noise by thresholding the empirical wavelet coefficients. The Sure Shrink threshold *t\** is defined by Eq. (4).

$$t^* = \min(t, \sigma\sqrt{2\log n}) \quad\quad\quad\quad\quad\quad (4)$$

Here, t denotes the value that minimizes SURE, $\sigma^2$ is the noise variance and n is the number of samples.

#### C. Bayes Shrink

Bayes Shrink was proposed by Chang, Yu and Vetter [11]. The purpose of this method is to minimize the Bayesian risk. It uses soft thresholding and is sub band-dependent where the thresholding is done at each band of resolution in the wavelet decomposition. It is also a smooth adaptive method as similar as the Sure Shrink procedure. The Bayes threshold is defined by Eq. (5).

$$t_b = \sigma_n^2 / \sigma_s^2 \quad\quad\quad\quad\quad\quad\quad\quad\quad\quad (5)$$

Here, $\sigma_n^2$ is noise variance and $\sigma_s^2$ is signal variance without noise. The definition of additive noise gives the following Eq. (6) and Eq. (7).

$$w(x,y) = s(x,y) + n(x,y) \quad\quad\quad\quad\quad (6)$$

Here, w(x,y) is noisy image, s(x,y) is original image and n(x,y) is added noise.

$\sigma_w^2 = \sigma_s^2 + \sigma_n^2$; where $\sigma_w^2$ is total variance.

$\sigma_n^2$ are computed as

$$\sigma_w^2 = \frac{1}{n}\sum_{x,y=1}^{n} w(x,y)^2, \sigma_n^2 = \frac{1}{n}\sum_{x,y=1}^{n} n(x,y) \quad\quad\quad (7)$$

The variance ($\sigma_s^2$) of the signal is computed as shown in Eq. (8).

$$\sigma_s = \sqrt{\max(\sigma_w^2 - \sigma_n^2, 0)} \quad\quad\quad (8)$$

### D. Neigh Shrink

This wavelet-domain image thresholding scheme was proposed by Bui and Chen [12] and it incorporates the neighboring coefficients. The shrinkage function for Neighshrink of any arbitrary window is expressed by Eq. (9).

$$\gamma_{i,j} = \left[1 - \frac{T_u^2}{S_{i,j}^2}\right] \quad\quad\quad (9)$$

Here, $T_u^2$ is universal threshold and $S_{i,j}^2$ is squared sum of wavelet coefficients in a given wavelet which is defined by Eq. (10).

$$S_{i,j}^2 = \sum_{n=j-1}^{j+1}\sum_{m=i-1}^{i+1} Y(m,n)^2 \quad\quad\quad (10)$$

The estimated center wavelet coefficient $\tilde{F}_{i,j}$ is calculated from noisy $Y_{i,j}$ as given in Eq. (11).

$$\tilde{F}_{i,j} = \gamma_{i,j} \cdot Y_{i,j} \quad\quad\quad (11)$$

Here, $Y_{i,j}$ = Noisy image within the given arbitrary window.

### E. Bilateral Filter

The Bilateral filter is a nonlinear filter proposed by Tomasi and Manduchi [13] and is used to reduce additive noise from images. Bilateral filtering technique smoothens the images while preserving edge, through nonlinear combination of nearby pixel values. The bilateral filter takes a weighted sum of the pixels in a local neighbourhood, which depend on both the spatial distance and the intensity distance. In this way the edges are well preserved and the noise is averaged out. Mathematically, the output of a bilateral filter at a pixel location x is calculated as shown in Eq. (12).

$$\tilde{I}(x) = \frac{1}{C}\sum_{y\in N(x)} e^{-\frac{(y-x)^2}{2\sigma_d^2}} e^{-\frac{(I(y)-I(x))^2}{2\sigma_r^2}} I(y) \quad\quad\quad (12)$$

Here, $\sigma_d$ and $\sigma_r$ are parameters controlling the fall-off of weights in spatial and intensity domains, respectively. $N(x)$ is a spatial neighbourhood of pixel $I(x)$, and C is the normalization constant. The parameters for bilateral filter are $\sigma_d = 1.8$, $\sigma_r = 2\sigma_n$ and windows of size 11x11 [13], [14].

### F. Multi Resolution Bilateral Filter (MRBF)

Multi resolution analysis has been proven to be an important tool for eliminating noise in signals. It is possible to distinguish between noise and image information better at one resolution level than another. Therefore, the bilateral filter is used for noise reduction in a multi-resolution framework [15]. An image is decomposed into its frequency sub-bands with wavelet decomposition. As it can be reconstructed, bilateral filtering is applied to the approximation sub-bands and wavelet thresholding to the detail sub-bands.

### G. Image matrices

There are various metrics used for evaluation of an image as mentioned in the introduction. Let the original noise-free image be $X(m,n)$, noisy image be $Y(m,n)$ and the filtered image be $\bar{X}(m,n)$, where m and n represent the discrete spatial coordinates of the digital images. Mean Square Error (MSE) and Root Mean Squared Error (RMSE) are defined as in equation (13).

$$MSE = \sum_{m=1}^{M}\sum_{n=1}^{N}(\bar{X}(m,n) - X(m,n))^2 \quad\quad\quad (13)$$
$$RMSE = \sqrt{MSE}$$

And Mean Absolute Error (MAE) is defined as in equation (14).

$$MAE = \sum_{m=1}^{M}\sum_{n=1}^{N} |(\bar{X}(m,n) - X(m,n))| \quad\quad\quad (14)$$

PSNR is defined in logarithmic scale, in dB. It is a ratio of peak signal power to noise power. Since the MSE represents the noise power and the peak signal power, the PSNR is defined as in Eq. (15).

$$PSNR = 10*\log_{10}\frac{L}{MSE} \quad\quad\quad (15)$$

UQI is derived by considering three different factors:

i. Loss of correlation

ii. Luminance distortion

iii. Contrast distortion.

It is defined as in Eq. (16).

$$UQI = \frac{\sigma_{f,\tilde{f}}}{\sigma_f \sigma_{\tilde{f}}} \cdot \frac{2\bar{f}\cdot\bar{\tilde{f}}}{(\bar{f})^2 + (\bar{\tilde{f}})^2} \cdot \frac{2\sigma_f \sigma_{\tilde{f}}}{(\sigma_f)^2 + (\sigma_{\tilde{f}})^2} \quad\quad\quad (16)$$

Where,

$$\bar{f} = \frac{1}{MN}\sum_{x=1}^{M}\sum_{y=1}^{N} f(x,y), \bar{\tilde{f}} = \frac{1}{MN}\sum_{x=1}^{M}\sum_{y=1}^{N}\tilde{f}(x,y)$$

$$\sigma_f^2 = \frac{1}{MN-1}\sum_{x=1}^{M}\sum_{y=1}^{N}(f(x,y) - \bar{f}),$$

$$\sigma_{\tilde{f}}^2 = \frac{1}{MN-1}\sum_{x=1}^{M}\sum_{y=1}^{N}(f(x,y) - \bar{\tilde{f}})$$

$$\sigma_{f,\tilde{f}}^2 = \frac{1}{MN-1}\sum_{x=1}^{M}\sum_{y=1}^{N}(f(x,y) - \bar{f})(f(x,y) - \bar{\tilde{f}})$$

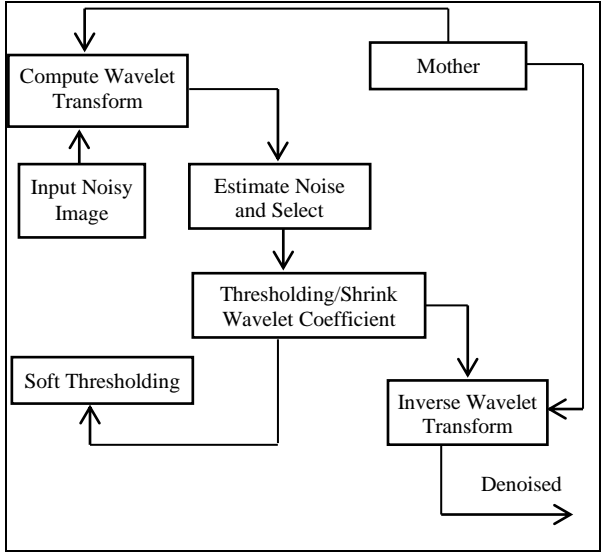

Fig. 2. Test-bed architecture for simulation.

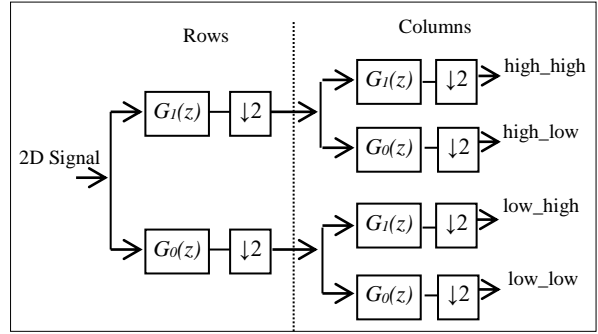

Fig. 3. 2D signal decomposition into various sub bands.

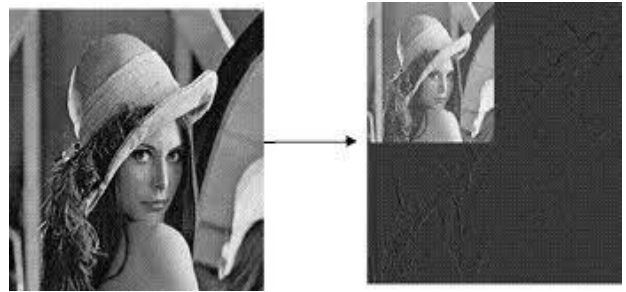

Fig. 4. First level decomposition of image Lenna.

## IV. METHODOLOGY

The work was carried out on six different images commonly used in filter design, which are Lenna, Goldhill, Boat, House, Peppers and Barbara. The experiments were conducted to observe the performance of different denoising methods quantitatively and visually using a MATLAB R2013a which provides various functions to carry out simulation. Overall procedure about the image denoising was simulated as shown in Figure 2.

Visushrink, Sureshrink, Bayeshrink, Neighshrink, Bilateral and MRBF were considered for denoising images. The test images were Lenna(512x512), Goldhill(512x512), Boat(256x256), House(512x512), Peppers(256x256) and Barbara(512x512).

### A. Addition of Noise

For comparative simulation, noisy images were created by adding Gaussian white noise (AWGN). The standard deviations of 10, 20, 30, 40, and 50 were used on our six standard test images. These noisy images were then denoised using several denoising techniques in wavelet domain and bilateral filtering.

### B. Computation of DWT

The discrete wavelet transform (DWT) is must for image denoising which requires the selection of different mother wavelet family such as Haar, Daubechies, Symlets, and Coiflets. The image can be transformed in wavelet domain using any of these wavelet family. In this paper, we have used the mother wavelet from Haar family.

### C. Decomposition of images into sub-bands

As stated in the previous section, the wavelet transform was first introduced for the time frequency analysis of transient continuous signals. It was then extended to the multi-resolution wavelet transform using Finite Impulse Response (FIR) filter approximation as demonstrated in Figure 3. After that, the first level decomposition on the noisy images was carried out. Figure 4 shows the process imposed on image Lenna.

### D. Thresholding and Threshold Estimation

The sub-bands $HH_k$, $HL_k$, $LH_k$ where k=1, 2, 3...j are called the details, where k is the scale, j denotes the largest or coarsest scale in decomposition, and $LL_k$ is the lowest resolution component.

All the Wavelet thresholding techniques viz. Visu, Sure, Bayes and Neigh were applied to the detail components of these sub bands to remove the unwanted coefficients that contribute to the noise. The inverse discrete wavelet transform was then applied to build back the modified image from its coefficients. In addition to those methods, the Bilateral and the MRBF filters, and the collaborative method (first Bayes and then Bilateral filter applied) were also used to process the noisy images in order to evaluate the individual performance.

## V. RESULTS ANALYSIS

For the performance analysis of thresholding and filtering techniques, different images were obtained using denoising methods and the image matrices were evaluated for those techniques. Based on the matrices, the performance of different image denoising techniques was evaluated.

Figure 5 presents the output of different denoising techniques at particular noise density for Image Lenna

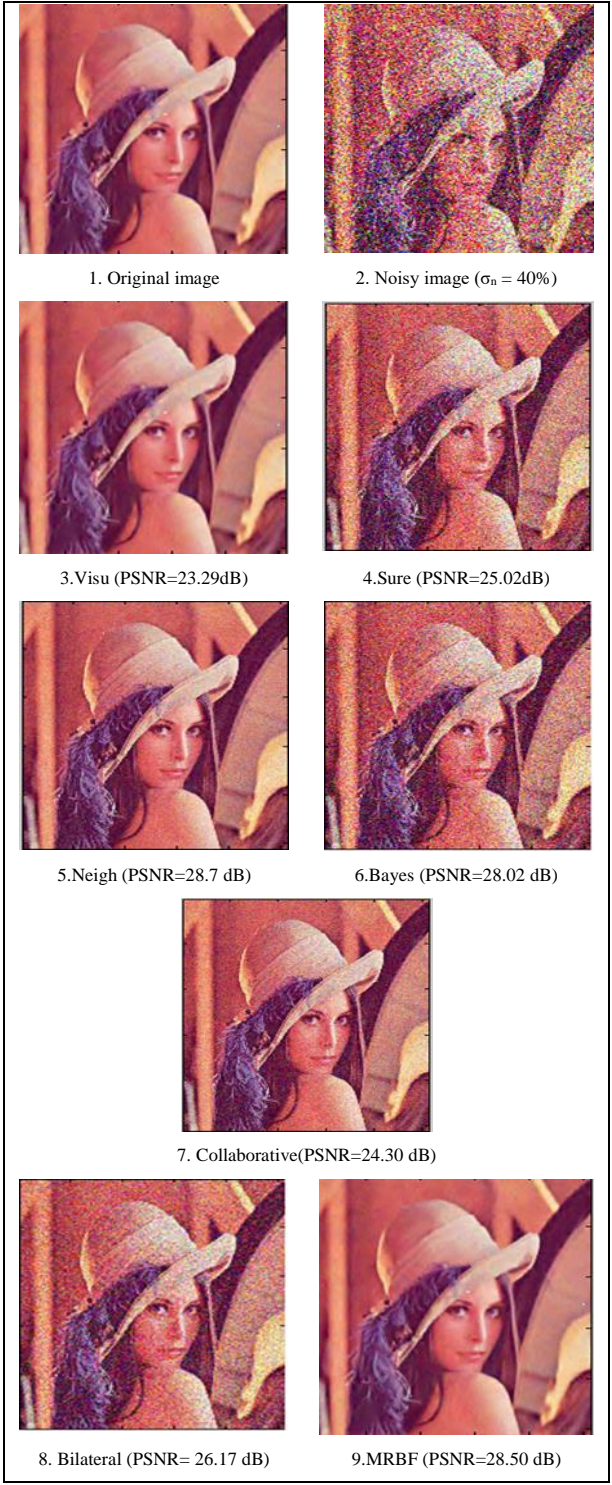

Fig. 5. Output of different denoising techniques at particular noise density for Image Lenna.

| TABLE I. | PERFORMANCE OF DIFFERENT FILTERING TECHNIQUE IN TERMS OF PSNR AT DIFFERENT AGWN |

| Peak Signal to Noise Ratio(PSNR) in dB | | | | | |
|---|---|---|---|---|---|
| | Standard Deviation of AGWN | | | | |
| **Filter type** | 10 | 20 | 30 | 40 | 50 |
| Visu | 26.89 | 25.74 | 24.73 | 23.29 | 21.82 |
| Sure | 25.41 | 25.32 | 25.17 | 25.02 | 24.77 |
| Bayes | 33.81 | 30.98 | 29.45 | 28.02 | 26.86 |
| Neigh | 35.82 | 31.98 | 29.45 | 28.7 | 26.86 |
| Bilateral | 31.07 | 29.10 | 26.75 | 26.17 | 24.48 |
| Collaborative | 29.30 | 28.78 | 25.6 | 24.30 | 21.10 |
| MRBF | 37.82 | 32.30 | 30.5 | 28.50 | 27.01 |

| TABLE II. | PERFORMANCE OF DIFFERENT FILTERING TECHNIQUE IN TERMS OF UQI AT DIFFERENT AGWN |

| Universal Quality Identifier (UQI) | | | | | |
|---|---|---|---|---|---|
| | Standard Deviation of AGWN | | | | |
| **Filter type** | 10 | 20 | 30 | 40 | 50 |
| Visu | 0.986 | 0.982 | 0.977 | 0.963 | 0.956 |
| Sure | 0.978 | 0.979 | 0.976 | 0.975 | 0.974 |
| Bayes | 0.998 | 0.985 | 0.984 | 0.971 | 0.968 |
| Neigh | 0.997 | 0.995 | 0.99 | 0.987 | 0.985 |
| Bilateral | 0.974 | 0.971 | 0.959 | 0.947 | 0.931 |
| Collaborative | 0.979 | 0.961 | 0.950 | 0.940 | 0.930 |
| MRBF | 0.998 | 0.989 | 0.986 | 0.978 | 0.973 |

(512x512). Table I and Table II were obtained for the image under various intensities of Gaussian noise. The result shows that Neighshrink is suitable for image denoising among other four wavelet transform filters along with Bayeshrink. The statistical data was visualized graphically as in Figure 6 and Figure 7 which were constructed from Table I and Table II respectively.

As shown in Table I, when σn=40, the PSNR values for Visushrink is 23.29 dB and that for Sureshrink is 25.02dB whereas the PSNR values for Bayeshrink and Neighshrink were observed above 28dB. This is because the estimated standard deviations in Visushrink and Sureshrink are more robust than the standard deviations of the sample. So the universal threshold tends to be higher for larger values of samples, eliminating many signal coefficients along with their noise. Hence Visushrink and Sureshrink do not adapt well to discontinuities and are less suitable denoising techniques as compared to Bayeshrink and Neighshrink.

Now to further analyze the obtained result, various types of images with different noise intensities were taken and applied with all our denoising techniques. Different result was obtained for six different images with various intensities of noise as shown in Table III. On average, the MRBF technique is 1.91 dB better than the original bilateral filter and 1.01 dB better than the Neighshrink wavelet thresholding while considering six different images with different AGWN intensity.

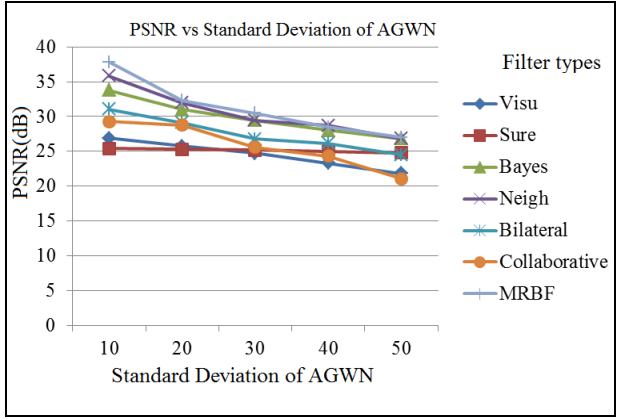

Fig. 6. Plot of Noise vs. PSNR for different AGWN intensities.

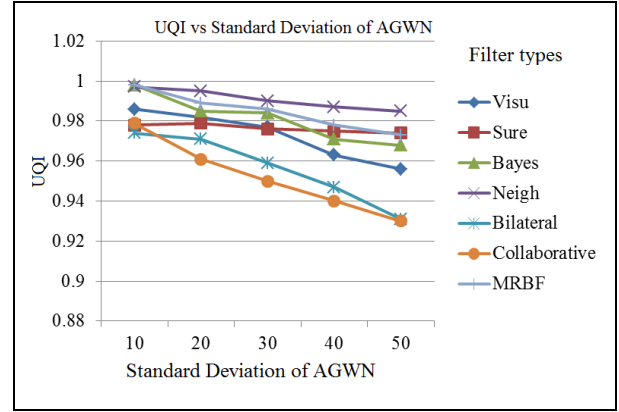

Fig. 7. Plot of Noise vs. UQI for different AGWN intensities.

TABLE III. PERFORMANCE OF DIFFERENT DENOISING METHODS FOR DIFFERENT IMAGES AT VARIOUS NOISE INTENSITIES

|  | $\sigma_n$ | VISU | SURE | Bayes | NEIGH | Bilateral | Collaborative | MRBF |
|---|---|---|---|---|---|---|---|---|
| Lenna (512*512) | 10 | 30.75 | 29.50 | 31.25 | 32.30 | 31.37 | 31.67 | 32.79 |
|  | 20 | 28.32 | 27.53 | 27.32 | 28.23 | 27.02 | 27.30 | 27.74 |
|  | 30 | 27.80 | 27.40 | 25.34 | 26,02 | 24.69 | 25.40 | 25.83 |
|  | 40 | 23.30 | 22.21 | 24.30 | 25.30 | 23.89 | 23.89 | 25.45 |
| Goldhill (512*512) | 10 | 30.56 | 30.08 | 31.94 | 31.89 | 31.93 | 31.60 | 32.98 |
|  | 20 | 29.50 | 27.69 | 28.69 | 30.30 | 28.80 | 29.56 | 32.48 |
|  | 30 | 27.23 | 26.13 | 27.13 | 28.90 | 27.50 | 27.56 | 29.50 |
|  | 40 | 23.33 | 22.34 | 23.67 | 24.56 | 23.40 | 23.90 | 24.56 |
| Boat (256*256) | 10 | 30.78 | 29.48 | 30.56 | 32.23 | 31,87 | 31.99 | 32.48 |
|  | 20 | 27.67 | 27.90 | 27.69 | 29.40 | 29.20 | 28.56 | 29.50 |
|  | 30 | 26.99 | 27.50 | 26.89 | 27.56 | 27.30 | 26.78 | 27.77 |
|  | 40 | 22.60 | 21.90 | 21.55 | 23.10 | 22.34 | 22.40 | 23.40 |
| House (512*512) | 10 | 31.12 | 31.77 | 33.07 | 33.34 | 33.01 | 31.40 | 33.67 |
|  | 20 | 27.10 | 27.90 | 27.12 | 29.90 | 28.10 | 28.90 | 30.20 |
|  | 30 | 26.23 | 27.13 | 26.13 | 27.99 | 26.07 | 27.34 | 28.13 |
|  | 40 | 23.10 | 23.60 | 23.23 | 23.50 | 22.67 | 23.01 | 23.80 |
| Peppers (256*256) | 10 | 31.93 | 31.67 | 31.49 | 33.32 | 31.89 | 31.23 | 34.62 |
|  | 20 | 29.30 | 29.80 | 27.85 | 29.88 | 28.01 | 29.34 | 29.24 |
|  | 30 | 27.02 | 27.02 | 25.73 | 26.67 | 26.07 | 26.89 | 31.37 |
|  | 40 | 23.23 | 23.45 | 23.12 | 24.60 | 23.09 | 23.87 | 24.78 |
| Barbara (516*516) | 10 | 31.98 | 31.39 | 33.27 | 33.70 | 33.39 | 31.67 | 34.48 |
|  | 20 | 29.23 | 29.54 | 30.27 | 31.20 | 30.29 | 29.08 | 31.28 |
|  | 30 | 27.45 | 28.54 | 28.60 | 29.10 | 28.62 | 27.34 | 29.33 |
|  | 40 | 23.40 | 23.67 | 23.78 | 24.56 | 23.67 | 22.90 | 25.56 |
| Average |  | 27.48 | 27.29 | 27.49 | 28.76 | 27.20 | 27.64 | 29.21 |

The collaborative method from Bayeshrink and bilateral filter applied together is found slightly better than just Bayeshrink and worse than just bilateral filter. Therefore, it is concluded that the improvement of the performance is not due to the combined effect of Bayeshrink and bilateral filter, but due to the multi resolution application of the bilateral filter when used with wavelet transform. Hence, MRBF is better filtering technique in terms of PSNR and UQI as compared with other wavelet transform techniques.

## VI. CONCLUSION

From the obtained simulated values, Neigh Shrink and Bayeshrink yield good performance under low variances of noise among different wavelet transform filtering techniques.

Bilateral filter is not so efficient in terms of PSNR and UQI but when it is applied with wavelet transform filter such as Bayeshrink, its PSNR and UQI get slightly increased. When bilateral filter with multilevel wavelet transform i.e. MRBF is used, its performance is increased significantly, with the multi-level bilateral filter, and the strong noise is eliminated most effectively. In multilevel bilateral filter, the improvement of the performance in terms of PSNR and UQI is not only due to the combined effect of Bayeshrink and bilateral filter, but due to the multi resolution application of the bilateral filter when used with wavelet transform. MRBF can use any type of wavelet thresholding method, such as Visu, Sure. Compared with Bayes and Neighshrink, although other methods with bilateral filter have good performance for the less noisy image, but when it comes to the strong noisy image, using same parameters with Bayeshrink has given the best output performance. However, Sure are Visu wavelet thresholding techniques are not effective for the noisy images because wavelet thresholding of these methods are based on the robust median estimation.

Since selection of the right denoising procedure plays a major role, it is important to experiment and compare the methods. In future, the work will be carried out on the further comparison of the other denoising techniques such as wavelet transform using Curvelets, Ridgelets and Fuzzy based wavelet shrinkage. If the features of the denoised signal were fed into a neural network pattern recognizer, then the rate of successful classification would determine the efficient denoising procedures. Besides, the complexity of the algorithms can be measured according to the CPU computing time flop. This can produce a time complexity standard for each algorithm. These two points can be considered as an extension to the present work.

## ACKNOWLEDGEMENT

The authors acknowledge the helpful discussion with Dr. Keshar Prasain.